# Integrators at War: Mediating in AI-assisted Resort-to-Force Decisions

Dennis Müller[1], Maurice Chiodo[2] & Mitja Sienknecht[3]

**Abstract:**

The integration of AI systems into the military domain is changing the way war-related decisions are made. It binds together three disparate groups of actors - developers, integrators, users - and creates a relationship between these groups and the machine, embedded in the (pre-)existing organisational and system structures. In this article, we focus on the important, but often neglected, group of integrators within such a sociotechnical system. In complex human-machine configurations, integrators carry responsibility for linking the disparate groups of developers and users in the political and military system. To act as the mediating group requires a deep understanding of the other groups' activities, perspectives and norms. We thus ask which challenges and shortcomings emerge from integrating AI systems into resort-to-force (RTF) decision-making processes, and how to address them. To answer this, we proceed in three steps. First, we conceptualise the relationship between different groups of actors and AI systems as a sociotechnical system. Second, we identify challenges within such systems for human-machine teaming in RTF decisions. We focus on challenges that arise a) from the technology itself, b) from the integrators' role in the sociotechnical system, c) from the human-machine interaction. Third, we provide policy recommendations to address these shortcomings when integrating AI systems into RTF decision-making structures.

[1] Centre for the Study of Existential Risk, University of Cambridge, Cambridge, United Kingdom. dm782@cantab.ac.uk
[2] Centre for the Study of Existential Risk, University of Cambridge, Cambridge, United Kingdom. mcc56@cam.ac.uk **[corresponding author]**
[3] European New School of Digital Studies, European University Viadrina, Frankfurt(Oder), Germany. sienknecht@europa-uni.de

## 1. Introduction: AI integrators as a hidden problem area

Integrating AI systems in resort-to-force (RTF) decision-making processes changes how war and peace are made. The integration of new technologies into existing decision-making structures is common in the military, but integrating AI-based systems constitutes a new situation. There has been comparatively little attention given to AI systems for high-level strategic military and political RTF decision-making (Erskine and Miller, 2024).[4] The technology itself, the different groups involved, and the human-machine interaction pose challenges to the proper functioning and responsible usage of AI in this morally sensitive context.

AI contributes to RTF decisions in many ways, from low-level data analysis, such as detecting enemy vehicles in an image (otherwise done by junior workers), to medium-level aggregation tasks, such as identifying patterns in enemy movements or communications (otherwise done by trained analysts), to high-level analysis, taking in a broad range of data and providing probable scenarios, perhaps alongside recommendations on potential courses of action (otherwise done by senior analysts). In this article we focus on the latter, where AI is carrying out work usually tasked to senior analysts and decision makers.

While research or public discourse about AI pitfalls often focuses on either those who develop[5] such AI systems (developers, companies, etc.), or the end-users[6] (political leaders, high-ranked military, etc.), that doesn't tell the whole story. There's another group that contributes to the (ir-)responsible functioning of the system, and that's integrators;[7] a group

---

[4] For example, only recently research on explainable AI in strategic decision making has begun; often with a focus on limited games, such as chess (e.g., Govaers, 2018). For RTF some attention has been paid to reducing risks and shortcomings of AI and general policy recommendations (e.g., Hoffman and Kim, 2023), potential user benefits (e.g., Void, 2024), and challenges to develop AI responsibly (e.g., Chiodo et al., 2024a), among other selected issues.

[5] Such as building safe AI, including "guaranteed safe AI" (Dalrymple et al., 2024), "restricted access" (Shevlane, 2022) or on "engineering pitfalls" (Morales-Forero et al., 2023).

[6] For example, existing research puts a lot of attention on using AI safely in everyday situations and lower to mid- level decision-making (e.g., Kushwaha, 2023), and thus, among others, on how (cognitive) biases play a role in their (safe) usage and in the development of explainable AI (Bertrand et al., 2022).

[7] Our distinction between developers, integrators, and users (Chiodo et al., 2024a) differs from that currently used in the regulation of civilian AI. For example, the EU AI Act focuses predominantly on providers and deployers of AI, talking about AI operators if they are one institutional entity. To put

often neglected in both academic and public discourse. They're responsible for integrating AI into already existing systems and organisational structures, thus having more responsibility than usual system engineers. They aim to avoid harmful effects both in the development process, e.g., so that developers do not lack information about use cases, and in AI usage, e.g., so that users understand what information is displayed by the AI and can properly interpret and query it. In this sense, integrators take a sandwich position between the developers and the users; they are responsible for balancing and translating between both groups and the machine. This article asks which challenges and shortcomings emerge from the integration process for AI in RTF decisions, and how to address them.

To answer this, we conceptualise the relationship between humans and machines on the one hand, and between different groups of actors relevant to integration on the other, as socio-technical systems (Trist & Bamforth, 1951). We use this approach to identify challenges and shortcomings in socio-technical relations that relate to the role of integrators, whom we distinguish from traditional systems engineers who act as the "maestro" coordinating (purely) technical expertise to build functioning systems (Ryschkewitsch et al., 2009, p. 83). The latter are typically positioned between engineering teams rather than between developers and military or political decision makers.[8] Integrators of AI in RTF typically go beyond the system engineers' tasks.

Of the three groups involved in the rollout of AI technology, the role of the integrator is invisible to most, provided they're doing their job well. This neglect, or rather ignorance, of integrators, leads to an overemphasis on the role of developers and users[9] when things go

---

more focus on the people involved, we decided to use developers and integrators instead. See Chiodo et al. (2024b) for a discussion of the relevant actors and regulatory approaches for the European civilian market.

[8] Though senior systems engineers may sometimes manage entire projects, overlapping with integrator responsibilities.

[9] For example, the summary of the North Atlantic Treaty Organization's (NATO) Artificial Intelligence Strategy consistently talks about developing and using AI, but only mentions integration when it focuses on the interoperability between different tasks (NATO, 2021). However, to meet NATO's six "Principles of Responsible Use of Artificial Intelligence in Defence", i.e., lawfulness, responsibility and accountability, explainability and traceability, reliability, governability, bias mitigation (ibid.), it's necessary to avoid merging the different groups involved in the creation of such systems.

wrong as at that point it's common to blame either "design faults" or "user error". Therefore, we argue to shift the focus to the integration of AI into RTF decision-making processes, and on the role of the integrators in this morally highly sensitive context of potential war. Integrators require a deep understanding of the activities, perspectives, and norms of both developers *and* users, who may have little understanding of each other. They have a distinct role in ensuring the proper functioning of any AI-based decision support system.

While the science and technology studies (STS) literature mainly focuses on discussing ways to increase the efficiency of socio-technical systems, we use it to identify specific challenges for the integrators of AI RTF systems. In this way, we approach the integration of AI into decisions about war and peace - and thus about life and death - from a very technical perspective. This seemingly bloodless approach is crucial to emphasise the various, sometimes conflicting, relationships between the groups of actors set into relation by the use of AI, and to make recommendations addressing the shortcomings, and for improving its integration process. With this focus, our article deals with complication 4 of Erskine and Miller's '*four complications*', on changes in organisational decision-making as the result of the integration of AI-enabled systems in state-level RTF decision making (see Erskine & Miller, 2024, pp. 139-40).

In what follows, we first discuss (in section 2) historical examples of the integration of new technologies into the military, highlighting what distinguishes previous technologies from the integration of AI systems into military decision making. Section 3 further conceptualises the relationships between humans, AI-enabled technology and the organisational context as a socio-technical system. Section 4 details some potential challenges arising from this configuration. And section 5 adapts the recommendations for developers by Chiodo et al. (2024a) into pillars of responsible AI integration (see Table 1). These challenges and pillars then form the basis of policy recommendations for the education and employment of integrators.

## 2. Historical examples of integrating new technology in the military

Integrating new technologies into military structures is no new phenomenon; they've repeatedly transformed warfare over time, making modern battlefields vastly different from those of the past. Technological development has always been driven by desires to improve one's capabilities against adversaries, i.e., by being faster, more accurate and, ideally, more successful (for a critical discussion of success in the context of war, see Renic, *in preparation*). Technology "has changed warfare more than any other variable [including] politics, economics, ideology, culture, strategy, tactics, leadership, philosophy, psychology" (Roland 2016, p. 1). However, each technological development requires careful integration to ensure that it enhances rather than complicates military processes and operations, thus making the group of integrators central for success. In the following, we illustrate the integration of new technologies and resulting challenges with the examples of aircraft carriers and cryptographic tools.

The integration of new technology is often complicated. Consider, for example, the integration of aviation into the navy. Warships have existed for millennia, yet significant time and effort was needed to integrate aeroplanes into the navy in the form of aircraft carriers. Even though fixed-wing aircraft were invented in 1903 (National Air and Space Museum, 2024), it took until 1910 for one to take off from a stationary ship (Moore, 1981), and until 1917 for one to land on a moving ship[10] (Iredale, 2015, pp. 20-21). Indeed, the first aircraft carriers were literally that: carriers of aircraft (seaplanes), which they would lower into and raise out of the water using a crane (Toppan, 2001). It took additional innovations, such as *arrestor gears*, to make their proper integration possible.[11] This is the work of engineers acting as integrators; making things that already exist work in harmony - in this case, ships, aircraft, and rope - while dealing with many groups, from aircraft engineers to aviators to naval captains, all of whom had different demands, requirements, and knowledge bases. But when done correctly, such integration changed the face of war. Aircraft carriers became the backbone of naval superpowers during the Second World War, and have remained so ever since.

---

[10] A task so hard that the first pilot to achieve it died five days later attempting the same landing.
[11] Landing a plane usually requires a long runway providing it sufficient space to halt using its own breaks. However, aircraft carriers have very short runways. So integrators invented arrestor gears, which were ropes (attached to weights) that a landing aircraft catches on to, rapidly slowing down the aircraft over a much shorter distance.

While the invention of aircraft carriers was a military success story, some innovations more clearly demonstrate their harmful effects when not integrated properly. Cryptography is such an example, which is by no means as visible as an aircraft carrier, but on the contrary is and remains invisible at best. Cryptography cannot be seen in action, takes minimal physical form, and plays no direct part in enabling physical force. It's an information tool, helping create and maintain a knowledge asymmetry favourable to those using it. Cryptographic tools are used right up to the highest levels of command, enabling all parts to carry out their communications in perfect secrecy, or so they hope. And while the mathematics that underpins most types of cryptography is extremely robust, and thus at the development stage such tools are often deemed "safe and secure", it is the human-computer interface that introduces most cryptographic weaknesses and vulnerabilities.

For example, poor use of cryptography, such as repeated use of location codewords, enabled American cryptographers in 1942 to determine that the Japanese were planning a naval attack on Midway[12] (Smith, 2001, p. 138). A better integration of the cryptographic tools, particularly a better education of the users, would have led to different protocols, such as using multiple different codewords. This was key to Admiral Yamamoto's defeat at the Battle of Midway, referred to as "one of the most consequential naval engagements in world history" (Symonds, 2018, p. 293). Twelve months later, over-reliance on cryptographic security meant the Japanese transmitted the entire escort itinerary of a flight transport for

---

[12] By broadcasting a fake message about a lack of water on Midway, the Americans tricked the Japanese into re-using their codeword AF to say "AF is running low on water", confirming AF was indeed Midway.

Yamamoto which the Americans decrypted, enabling them to locate, intercept, and shoot down the transport, killing Yamamoto (McNaughton, 2006, p. 185). Here, better integration would have helped users understand to practise principles of data minimisation: transmit as little as possible. Inadequate integration of cryptography and its protocols cost Yamamoto the Battle of Midway in 1942, and his life in 1943.

In both scenarios, integrators played a decisive role: in scenario one for the better, and in scenario two for the worse. Socio-technical systems such as cryptography are fraught with difficulty, precisely at the human-machine interface where they most often fail. Ensuring that the interface functions effectively is the responsibility of the integrators. Getting a plane onto a moving ship is surprisingly difficult; getting a secure message onto one is even harder. So what can we learn from these examples about the integration of AI-based systems into RTF decision-making?

## 3. The integration of AI-enabled systems into the military and the constitution of a socio-technical system

One might argue that integrating AI is almost identical to integrating cryptography. This is not true. One key difference is that users of cryptography know what they want (transmit messages securely) and failure is falsifiable (enemies read the message and act on it); while users of AI in RTF decision making often do not know *a priori* precisely what they want the tool to do, beyond "give the right advice or analysis", and it's unfalsifiable when it fails; "what could have been" when reviewing a decision to start a war is impossible to know. Think about basing a preemptive strike on the AI-interpretation of adversarial troop movements; this is near-impossible to falsify. At all points, information will never be perfect, so neither will decisions. Hence, the goal for integrators in AI-RFT is nowhere near as well-defined as in cryptography. Perfection, as understood in the cryptographic setting, is an elusive target here. So what should integrators be aiming for, and how should they hope to achieve it?

Integrating AI in RTF decision making presents genuinely new challenges not faced by those integrating other (military) technologies. Contrasting technology used to achieve a *specified* goal, AI integration leads to a different composition of relationships between actors and technology, because AI systems act (semi-)independently (e.g., basing recommendations or predictions on independently identified patterns in vast datasets). Such systems may heavily influence top-level decisions on whether to start a war, and even if AI contributes to a well-reasoned RTF, there will undoubtedly be human injuries and loss of life on both sides, military and civilian. The decision to go to war is never bloodless, even if it is "right". And the integration of such systems varies, depending on their degree of automation; they're either intended to replace a human decision maker, or assist in decision-making (see Vold, *in preparation*).

Integrating weapons technologies doesn't happen in a normative vacuum. Indeed, it's constrained by international guidelines. This also applies to AI. In the 1977 Additional Protocol to the Geneva Conventions, Article 36 stipulates all new weapons are subject to the Geneva Convention. Arguably, the development, integration and use of AI RTF decision systems must follow the same rules as other weapons, means and methods of warfare, requiring signatories to determine if the "study, development, acquisition or adoption of a new weapon, means or method of warfare [...] [and] its employment would, in some or all circumstances, be prohibited by this Protocol or by any other [applicable] rule of international law." This is tricky for AI RTF decision-systems, as existing protocols and international law are falling behind on this. "To determine whether its employment would, in some or all circumstances, be prohibited" cannot be answered conclusively, as the functionality and operations of AI systems cannot necessarily be (pre)-determined. The fact that the technology is so advanced, yet so unaccountable (e.g., Erskine, 2024; Sienknecht, 2024), poses particular challenges, since its effects cannot be fully predicted.

However, integrating new technologies into organisational settings restructures existing, and creates new, actor-technology relations. It creates a (de facto new) socio-technical system (Emery & Trist, 1969), incorporating both existing elements from the "old" organisational and system setting and aspects which are unique to the new technology, in complex ways. Research on socio-technical systems originated in the 1950s from the work of Eric L. Trist on the UK coal mining industry. It focuses on 1) integrating new technologies into social systems, 2) human-machine-interaction, and 3) restructuring of organisational structures (Karafyllis, 2019, p. 300). But how does such a socio-technical system constitute through the integration process of AI into military decision-making structures?

Context-specific factors make each integration process, and thus each socio-technical system, unique. Nonetheless, we can sketch a common integration process which might help to better understand the constitution of such socio-technical systems. Each integration of RTF AI, though unique, follows a common trajectory. It begins with discussions between political and military stakeholders about technological solutions to specific problems. Ideally, integrators are involved early to manage expectations (e.g., feasibility of deploying AI in resource-limited locations) and provide developers with essential background knowledge for creating appropriate models. During this initialisation phase, developers, users, and integrators collaborate to define the AI system's specifications. So, in this first step, all three actor groups are set in relation to each other via specifying the details of the AI system to be built.

In the development and data collection phase, close cooperation between users, integrators, developers, and Intelligence Services ensures the model is trained on relevant, adequate, and well-curated data. Given the high stakes of RTF decisions, the model must undergo rigorous testing before deployment, incorporating feedback and updates for further refinement. Once validated, the model gets integrated into the decision-making structure. Integrators are responsible for this, and must ensure it complements other intelligence sources and (ideally) aligns with existing processes. This step is crucial in determining whether AI can be used without creating flaws in the decision-making process and whether AI information complements existing information obtained by analysts and intelligence. At this point, users are trained in the functionalities of the AI system and how to interpret the results, and user feedback helps refine the system further.

Once the AI is successfully (technically) integrated, the socio-technical system operates (semi-)independently, necessitating ongoing maintenance, updates, and emergency strategies for potential system shutdowns. Such strategies are vital to address flaws or prevent cascading failures, especially in security-critical contexts where the technology interacts with multiple processes.

This broad description of the integration process underlines that the socio-technical system's constitution starts long before the AI is ready for use. From planning, to development, integration, and usage, the socio-technical system changes and reconfigures. However, this description is a best-practice example of AI integration leading to a robust socio-technical system. But research has revealed several problems in the composition of the socio-technical system regarding the design, usage, and organisational restructuring (Pasmore et al., 2019, p. 68). One decisive problem is the interaction of social and technical factors: focusing on only one can lead to unpredictable and detrimental relationships within the system (Walker et al., 2008). The "interaction of social and technical factors creates the conditions for successful (or unsuccessful) system performance" (ibid., p. 480). In the present case, it's both about interaction between the social groups involved, and between the groups and the machine. Given the complexity of AI, the different logics of the involved groups, and the necessity to tie them together, we can expect multiple challenges with serious consequences in the context of potential war.

**4. Challenges arising in AI-based socio-technical systems**

Integrating AI systems into high-level RTF decision making bears several challenges that emerge from the specific context in which the AI system is embedded. While military decision processes for use-of-force are by now well-understood, well-rehearsed and continuously improved (see Center for Army Lessons Learned, 2024), decision-making processes on the RTF on the other hand are riddled with more known and unknown unknowns, including: i) each decision to go to war typically involves a new enemy, ii) there are fewer historical decisions available to train the AI, i.e., there's only one decision to go to war vs. many decisions to use force in each war, and iii) historical data may be inappropriate in the new context. In short: developers have to build an AI with insufficient available data, and integrators have to put it into a decision chain with a static structure but whose context is dynamic and unique each time. In military lingo: deciding on a mission is different from planning and executing said mission, and it's more difficult to practise, rehearse, and test. This presents new, unique challenges for everyone involved, which may not be addressable by the two dominant trends of data- and model-centric AI (compare Zha et al., 2023). For the former, we lack sufficient data. For the latter, we lack (among others) sufficient understanding to (adequately) mathematically model the involved decision spaces and thought processes, thus limiting approaches that directly integrate human knowledge into the AI training instead of relying on historical data sets (see Deng et al. (2020) for a discussion on integrating human knowledge into AI).

*4.1 Challenges arising from the technology itself*

As mentioned earlier, AI differs from technologies that are mere tools in the hands of humans, especially so-called "3rd wave AI technology". The United States Government Accountability Office (March 2018) describes three historical waves of AI development: 1) expert systems using logical reasoning; 2) classical statistical learning; and 3) "contextual adaptation", i.e., building on the previous waves to develop AI systems capable of "contextual sophistication, abstraction, and explanation" (Ibid., p. 18.). This third wave is ongoing and encompasses current attempts to build and integrate AI for RTF decision making.

Integrating 3rd wave AI is the most difficult one. 1st wave AI systems were often easier to explain and interpret due to their limited reasoning capabilities and use of hard-coded logic. 2nd wave AI systems were still rather limited in scope, requiring limited contextual understanding (e.g., translation of documents or identification of targets). Integrating 3rd wave AI systems requires extensive contextual domain knowledge on multiple fronts. Thus, those building such systems have been identified as potential bottlenecks for future development and improvements (Schuering and Schmid, 2024, p. 251), making their integration more demanding.

The AI-based technology to be implemented into military decision-making processes preconditions the range of possibilities in such processes themselves. The technological (Winner, 1980) and mathematical (Müller & Chiodo, 2023) aspects of these AI systems thus have politics. One must consider this to responsibly integrate AI into existing and new organisational structures. These factors range from the lack of creativity of the AI, as it bases its decisions on the past, to the missing falsifiability of its decisions.

In socio-technical systems, it's necessary to consider that "optimisation of either socio, or far more commonly the technical, tends to increase not only the quantity of unpredictable, 'un-designed', non-linear relationships, but those relationships that are injurious to the system's performance" (Walker et al. 2008, p. 480). Hence, successful integration of AI into RTF decision-making processes requires simultaneous optimisation of humans, machines, and organisational structures (for further discussion of military decision-making systems as complex adaptive systems (CAS), see Osoba, 2024, *in preparation*). At present it seems the military is enjoying a boost in the technological dimension (with the advancement of AI), without a similar boost in the socio-component (insufficient training of those working with AI), leaving the human-machine teaming rather unbalanced: powerful AI-enabled systems, poorly educated humans. Similar examples arose in the intelligence branches post-9/11 where new demands on analysts and their increasing number, required new forms of training and led to an (ongoing) debate about their professionalisation (Marrin, 2009; Gentry, 2016).

*4.2 Challenges arising from the role of integrators*

Integrators link two largely disparate groups: 1) AI developers: technically trained but often with little, if any, training or experience in politics, international relations, or military activity, and 2) AI users: mainly trained in humanities, social sciences or law, and come in a variety of positions, including politicians, political and intelligence analysts, and high-ranking military personnel, often with little, if any, technical training.[13] Senior military officials seldom understand deeper technical aspects of AI, and developers often lack expertise in security studies or the normative and legal restrictions on RTF. Yet integrators must somehow understand these two fields of expertise.

Within this demanding socio-technological context, the role of integrators further increases. However, they're often ignored as relevant actors, neither identified as those primarily building nor using the technology. As long as integrators do their job well, they're invisible to everyone. Admittedly, identifying integrators is often difficult given their varying forms of employment and positions within differing institutional structures, e.g., in the US military, this role appears to be primarily spearheaded by the Chief Digital and Artificial Intelligence Office, which subsumed the former Joint Artificial Intelligence Center in 2022. Different branches of the intelligence services and military also have their own developers, integrators and users; the existence of high-level AI offices doesn't necessarily imply the existence of a centralised oversight agency.

Integrators are generally "literate in data terminology and processes, especially in the military context, and use this knowledge alongside their operational intelligence experience to bridge the gap between industry or research institutions, and the Field Army".[14] As the link between developers and users, integrators bear a great sociological burden. They require a deep understanding of the activities, perspectives, and norms of both developers *and* users, who may have little understanding of each other. As described earlier, integrators should be involved from the onset, even before the model is built, to facilitate an uncomplicated integration. However, the positions of integrators in the wider work context and the organisation's structures may lead to power imbalances between the three groups. This asymmetry in hierarchy can, among other things, lead to employee silence during the initialisation, integration, usage and maintenance of the AI. This is extremely detrimental for integrators of high-stakes AI systems as they must be able to address problems, worries, and risks openly with users, their team, and developers.

---

[13] Traditionally, many analysts have a non-technical background and are trained to understand (groups of) people.

Furthermore, the limited number of users, alongside potential power imbalances between them and the integrators, can severely limit available testing options. Classical testing regimes, such as A/B testing user interfaces,[15] may be largely ineffective with only a handful of potential powerful users at any time, and lack of continuity between periods of office. Additionally, if the integrators disagree with the ideology and politics of users like politicians or high-ranking military officials, such users could overwrite current or past integrators' decisions, or worse, fire those employed as external consultants. Ultimately, integrators' expertise may be trumped by user power and political will. Examples of this exist, such as the unwanted purchase of Abrams tanks by the US Congress when the Army would have preferred different tools for their arsenal (see Sisk, December 18, 2014).

These power imbalances and (workplace) politics can negatively affect the options available to integrators and thus their morale to "go the extra mile" when integrating AI responsibly. Morale is a well-known, essential aspect in the military and for software developers (e.g., Besker et al., 2020; Whitaker, 1997). However, it's also crucial for those who must perform the difficult task of integrating AI in stressful, adversarial circumstances. The saying "from an engineer's perspective the users are the problem, and from a user's perspective the engineers are the problem", manifests in unique ways when the users have more (institutional) power.

---

[14] Biographical note provided by a participant at a recent workshop on the use of AI-assisted decision making in the military who is a Military Intelligence Data Specialist and wished to remain anonymous.
[15] See Quin et al. (2024) for an overview of A/B testing.

Likewise, the specific expertise needed to successfully integrate AI at the nexus of the political and military system might increase the power of the integrator. Just like other technical areas, integrators' (political) bargaining power mostly comes from their creation bottleneck in socio-technical systems; without integrators, the system can't be implemented. However, it's a largely open question if and how they could (and should) use this to change the course of RTF decision systems.

*4.3 Challenges arising from human-machine interaction*

We discussed how integrators are stuck in a sandwich position between developers and users. To perform an ethically aware and responsible integration of AI, integrators must look in two separate directions: they must understand the users and their (mis)behaviour, but also their suppliers (i.e., the AI developers), and align the expectations of the users with the technical feasibilities and capabilities. A developer-user mismatch jeopardises safe AI integration. Users may try to game the AI or misappropriate it, even if developed responsibly with limited use cases. Similarly, developers may have overlooked certain aspects of the problem, user expectations, or limitations in what users understand or can do.

For example, users may prefer simple AI inputs and outputs, as demonstrated in certain use-of-force and assisted planning decision systems by the US military in 2003: "[users] consistently looked for and asked for one simple way to enter the statement, and shied away from the rich, flexible, but necessarily complex approach the tool offered" (Forbus et al., 2023, p. 24). Such users may prefer gist-based reasoning over analytical verbatim-based reasoning (compare Reyna, 2024), thereby introducing biases into risk perception and decision making. As developers are looking for the "political gist" necessary to develop the system, users are looking for the "technical gist" necessary to interpret the results. Thus, Bosch and Bronkhorst (2018, p. 1) suggest that AI for military decision making must "adapt itself dynamically to the decision maker, by taking into account his objectives, preferences, and track record (e.g., susceptibility to bias)", necessitating *early* user involvement in development and integration.

This pressures integrators, who must balance user preferences, technical options, and contextual necessities. For AI, differing preferences can be particularly tricky, ranging from simple interface preferences and information presentation, to higher-level epistemic issues of differing assumptions about the representative nature and rationality of AI, its outputs and its mathematical formulation, including (social) justice issues. Across wider mathematics, and specifically AI, we see a chasm between those viewing mathematics and AI as neutral and pure, and those understanding its culture- and context-dependency (see also Rittberg, 2024; Müller et al., 2022); users, integrators and developers may likely take differing positions here, depending on their training and experience with mathematics and technology. They may not necessarily be aware of these biases and the standpoints of others.

Balancing different needs is especially difficult for 3rd wave AI systems, as the technical options available for development are still comparatively limited, and the requirements, in particular those concerning context and analysis, are ever-increasing. Further complications arise as existing research doesn't always address practical limitations, e.g., research on explainable AI often focuses on users with good technical knowledge (Saeed & Omlin, 2023, p. 7). This isn't necessarily true, even for well-educated decision-makers as users.

These are some core challenges emerging from integrating AI into RTF decision making and the constitution of a sociotechnical system. In the next section, we outline how such integration can be carried out more responsibly, by introducing the "10 pillars of responsible integration".

## 5. How to address challenges in the integration process?

Having discussed challenges arising from this human-machine interaction, we now give "10 pillars of responsible integration" that address different challenges that might arise during the integration process of an AI system into RTF decision making. It spans from initialisation concerns, over data handling and feedback loops to emergency response strategies, and thus addresses the full life-cycle of AI integration (see Table 1 for more details):

1. Initialisation.
2. Diversity and perspectives.
3. Handling data and development.
4. Data manipulation and output inference.
5. Interpretation of the problem and its technical solution.
6. Communication and documentation.
7. Falsifiability and feedback loops.
8. Explainability and safe AI.
9. Technical artefacts and processes have politics.
10. Emergency response strategies.

These 10 pillars draw on established tech-ethics frameworks, follow the AI lifecycle, build on decades of experience working with mathematicians and other technical experts, form part of a long project on ethics in mathematics, and have been tried and tested in industry, research and teaching (see Chiodo et al. 2024a; Chiodo & Müller, 2023, pp. 5-6). As happens for all mathematical work (Chiodo & Müller, 2023), the early and late pillars are less technical than the middle, reflecting the multidimensional challenges faced by integrators discussed earlier. Integrators must be experts of socio-technical systems, so their pillars go beyond technical aspects present for developers (Chiodo et al., 2024a) and general aspects present in all mathematical work (Chiodo & Müller, 2023); they also deeply connect with the (uniquely powerful) users. We argue that due to the complexity and manifold challenges in the AI integration process, all three groups - developers, integrators, users - must address these pillars to avoid harmful AI-assisted RTF decision systems, albeit with different tasks. Table 1 summarises the likely mismatches and problematic areas in the sociotechnical setting described above, based on the knowledge gaps of each group, and their differences in worldviews and workplace culture expectations. Notably, the pillars where developers are strong, users are weak, and vice-versa.

Thus, our examination of socio-technical systems reveals that, in line with technological advancement, the social aspect - encompassing those involved in human-machine interaction - must undergo a corresponding evolution. The potential challenges emerging during the constitutive process highlight that a pivotal requirement for productive human-machine interaction is the **education of all involved parties**. AI for strategic decision making is still in its infancy, both in military and corporate settings (Stone et al., 2020). Given the limited research in this area, there's also negligible education available for those wanting or obliged to do this. Such training is notably absent from many university curricula, resulting in a lack of education among engineers, computer scientists, or other technical experts hired as integrators. These individuals likely lack the requisite knowledge of complexities, risks, and issues involved. It's an unknowable unknown for educators, as the subject remains in its infancy, and curricular content yet to be determined. Overcoming this knowledge deficit may necessitate closer military-civilian research collaboration. It's therefore evident that **funding and supporting research into AI integration for strategic decision-making** is necessary to increase knowledge about potential benefits and drawbacks. Given these limitations, we must **develop specific recommendations on hiring integrators** to keep the military's capabilities up-to-date, and morally and legally aligned.

Despite uncertainty regarding specific challenges emerging from AI integration, we've identified several associated with human-machine interaction, the sandwich position of integrators, and the technology itself. One is **enhancing relationships between the three groups**. It's therefore crucial to acknowledge integrators' significant role in the rollout of new technologies; they aren't junior technical positions, rather, they're complex, safety-critical roles. Such considerations should be reflected in integrators standing among developers and users, the competencies attributed to them, and the contract conditions. This would **enable the resolution of power imbalances** arising in such situations, particularly when integrators are (temporarily) contracted, or when their work is modified or overruled.

Educating integrators along the 10 pillars for responsible AI integration will help address these challenges. We recommend **integrators receive explicit, mandatory induction training** reflecting their specific role, forthcoming challenges, and unique responsibilities. To further enhance their reputational standing among users, we recommend **implementing requirements for facilitated discussion between integrators and users** during the integration process and during long-term maintenance.[16] Each implementation domain is distinct, as are the senior military or government officials involved, and the AI itself. Therefore, adjustments are inevitable, and troubleshooting essential, throughout the process. Organisational structures should be adapted to the different challenges and requirements of integrating AI into the decision-making process (see Ryan, *in preparation*, on the necessity of military institutions learning and adapting to changing technologies).

It is essential to **facilitate discussions**, as integrators and users might speak very different professional languages. This supplements **establishing minimum standards for integrator responsibilities**. It is also essential to provide clear instructions and comprehensive documentation helping integrators find compromises between users' expectations regarding AI functionality and operation, and developers' (technical) constraints. This could be achieved through **establishing standards or codes of conduct**, including well-defined guidelines and rules of accountability. Ideally, these should be interoperable between allied nations, aligning with existing discussions on use-of-force decisions, such as command accountability (e.g., Kraska, 2021), and the legal interoperability between allies, including NATO (e.g., Hill & Marsan, 2018).

---

[16] See Hoffman and Kim (2023, p.2) who also suggest to "involve senior decision-makers as much as possible in development, testing, and evaluation processes for systems they will rely upon, as well as educating them on the strengths and flaws of AI so they can identify system failures".

The "10 pillars of responsible integration" represent an extended discussion about developers of AI-assisted RTF decision systems started by Chiodo et al. (2024a) to those integrating such systems. The pillars might serve as an orientation for the practical work of integrating AI into the RTF. By identifying potential challenges in the socio-technical systems and offering very practical solutions to them, we hope to contribute to facilitating the functional and responsible integration of AI into RTF decision-making processes - a development whose effects can only be surmised today. Of course, no amount of education or technological improvement can prevent the potential harm caused by an AI system deciding to go to war. Thus, society should do everything possible to address the potential challenges and complications arising from this new scenario.

**Table 1: Pillars for developers, integrators and users, colour coded by relative competences[17]**

These are the high-level pillars, with short 1-line descriptions, as presented in (Chiodo & Müller, 2023), with slight modifications for the "Integrators" and "Users" columns; the "developer" column matches (ibid., 2023) almost identically, replacing instances of "mathematics" with "AI", "resort-to-force AI decision-making", or some combination of these. There, these pillars are explored in much greater depth, with many more sub-questions, and sub-sub-questions. What is presented here is the "tip of the iceberg", laid out side-by-side for developers, integrators, and users, to help compare how the common issues manifest in all three groups.

We use the following demarcations within Table 1 to identify the relative competencies or shortcomings of each group in each pillar:

<mark style="background-color:green">Normal text, green</mark> = well versed, highly competent, able to deal with most unknown problems and new situations.

<mark style="background-color:yellow">**Bold text, yellow**</mark> = general awareness with a reasoned and structured approach to the typical known problems.

<mark style="background-color:red">**ALL CAPITALS AND BOLD TEXT, RED**</mark> = general lack of awareness and understanding, only able to deal with very simple situations and problems.

| Developers | Integrators | Users |
|---|---|---|
| **1. INITIALISATION:** Should you be providing AI for this resort-to-force system? Do you understand the AI's overall purpose? | **1. Initialisation:** Can you responsibly integrate this AI product or service, and should you even do so? | **1. Initialisation:** Why are you making use of this AI in your RTF decision-making, and should you even do so? |

---

[17] Based on the authors' experience working with and consulting mathematicians and AI experts on responsible development and integration. See also (Chiodo & Müller, 2023; Chiodo et al. 2024a) and the Ethics in Mathematics Project (www.ethics-in-mathematics.com).

| 2. DIVERSITY AND PERSPECTIVES: Do you and your co-workers have sufficient perspective of resort-to-force AI decision-making, and do you understand your limitations and biases? | 2. Diversity and perspectives: Do your co-workers, superiors, suppliers, and you have sufficient perspective of resort-to-force AI decision-making, and do you understand the limitations and biases in your thinking? | 2. Diversity and perspectives: Do you have sufficient perspective in your team/group to understand and question what this AI is doing/suggesting, and how to best use it? Do you understand how to reconcile your and the AI's biases and style of thinking? |
|---|---|---|
| 3. Handling data: Are you responsibly using authorised and morally obtained datasets for your AI training? | 3. Handling data and development: Are you and your suppliers using authorised and morally obtained datasets and their byproducts, responsibly and securely? | 3. HANDLING OUTPUTS: Are you and/or the people in your decision-making circle able to understand the outputs and suggestions this AI is providing, at a sufficient level for your situation. |
| 4. Data manipulation and inference: Do you have the expertise to properly manipulate data ensuring quality, ethics, and its relevance to resort-to-force situations? | 4. Data manipulation and output inference: Do you and your suppliers have the expertise to properly manipulate data ensuring quality and ethics, and can your users properly handle system outputs? | 4. OUTPUT MANIPULATION AND INFERENCE: Are you able to make effective and appropriate use of the outputs this AI is providing, in your decision-making and action-taking? |
| 5. Technical interpretation of the problem: What optimisation objectives have you chosen, and | 5. Interpretation of the problem and its technical solution: What type of artificial | 5. INTERPRETATION OF THE PROBLEM: Do you understand what the AI system is trying to do |

| what are their real-life consequences? Who else is impacted by this? | intelligence, optimisation objectives and constraints have you chosen, and what are their real-life consequences? Who might the other impacted parties be? | and to tell you, and how that (mis)matches with what you want to achieve? |
|---|---|---|
| **6. Communication and documentation:** How are you commenting and documenting your AI development and functionality? Are you communicating that to all relevant parties - in particular integrators and users? | **6. Communication and documentation:** Are you properly considering how to comment and document your work and communicate the results to those who need them? | **6. Communication and documentation:** Are you documenting how the AI outputs have been factored into your decision-making? Do you have an effective way to communicate to the populace how AI assisted your decisions, in a way they will understand and accept? |
| **7. FALSIFIABILITY AND FEEDBACK LOOPS:** Are your AI outputs sufficiently falsifiable for resort-to-force decision-making? How are you dealing with feedback loops from retraining on data impacted by your AI? | **7. Falsifiability and feedback loops:** Is your work falsifiable, and can you handle its large-scale impact and any feedback loops that arise? | **7. Falsifiability and feedback loops:** Are you using this AI system in a context or domain where its outputs, and your decisions, are falsifiable? Can you handle its large-scale impact and any feedback loops that arise? |

| 8. Explainability and safe AI: Are your AI's outputs interpretable and explainable to users? Are you conducting proper monitoring and maintenance of your AI's performance? | 8. Explainability and safe AI: Is your system integration explainable, interpretable and followed up with proper monitoring and maintenance? | 8. Explainability and safe AI: Is this system providing outputs to you that are explainable and interpretable by you? Has it been properly monitored and maintained leading up to its use? |
|---|---|---|
| 9. TECHNICAL ARTEFACTS HAVE POLITICS: What are the non-technical aspects of your AI's use, and the politics of its operation domain? How do you earn trust in yourself and your AI? | 9. Technical artefacts and processes have politics: Are you aware of other non-technical aspects and the political nature of your work? What do you do to earn trust in yourself, your system and its integration? | 9. Technical artefacts have politics: Are you aware of the political nature of using a/this AI in a resort-to-force decision-making context? Is it trustworthy to you, and to your citizens? |
| 10. EMERGENCY RESPONSE STRATEGIES: What is the non-technical response strategy in case your AI malfunctions or causes unintended consequences? Do you have anyone to support and help you deal with the fallout? | 10. EMERGENCY RESPONSE STRATEGIES: Do you have a non-technical response strategy for when things go wrong? Do you have a support network, including peers who support you and with whom you can talk freely? | 10. Emergency response strategies: How do you reconcile when the AI diverges or disagrees with an overwhelming consensus in your team, or with public perception? How does this AI fit into the decision-making of your supporters and allies? |